\documentclass[preprint,review,12pt]{elsarticle}

\usepackage{amssymb}

\usepackage{lineno}
\usepackage{amsmath}

\usepackage{amsmath}
\usepackage{algorithm}
\usepackage[noend]{algpseudocode}

\usepackage{subcaption}

\usepackage{hhline}

\usepackage{hyperref}

\usepackage{multirow}

\usepackage{booktabs}

\usepackage{geometry}
\usepackage{pdflscape}

\journal{Smart Agricultural Technology}

\begin{document}

\begin{frontmatter}

\title{Exploring Cluster Analysis in Nelore Cattle Visual Score Attribution}

\address[label1]{Universidade Católica Dom Bosco, Campo Grande, Brazil}
\address[label2]{Universidade Federal de Mato Grosso do Sul, Campo Grande, Brazil}
\address[label3]{Universidade Estadual de Mato Grosso do Sul, Campo Grande, Brazil}
\address[label4]{Kerow Soluções de Precisao, Campo Grande, Brazil}
\address[label5]{Embrapa Beef Cattle, Campo Grande, Brazil}

\author[label1]{Alexandre de Oliveira Bezerra}
\ead{alolbez@hotmail.com}

\author[label4]{Rodrigo Gonçalves Mateus}
\ead{rgmateus1.zoo@gmail.com}

\author[label3,label4]{Vanessa Ap. de Moraes Weber}
\ead{vanessaweber@uems.com}

\author[label4]{Fabricio de Lima Weber}
\ead{fabricio.weber@gmail.com}

\author[label1]{Yasmin Alves de Arruda}
\ead{yasmin\_alvesarruda@hotmail.com}

\author[label5]{Rodrigo da Costa Gomes}
\ead{rodrigo.gomes@embrapa.br}

\author[label1]{Gabriel Toshio Hirokawa Higa}
\ead{gabrieltoshio03@gmail.com}

\author[label1,label2]{Hemerson Pistori}
\ead{pistori@ucdb.br}
%

\begin{abstract}
Assessing the biotype of cattle through human visual inspection is a very common and important practice in precision cattle breeding. This paper presents the results of a correlation analysis between scores produced by humans for Nelore cattle and a variety of measurements that can be derived from images or other instruments. It also presents a study using the k-means algorithm to generate new ways of clustering a batch of cattle using the measurements that most correlate with the animal's body weight and visual scores.

\end{abstract}



\end{frontmatter}



\section{Introduction}
\label{intro}

Visual scores are widely utilized for biotype evaluation in beef cattle production, aiming at identifying individuals of better performance, precocity and with better distribution of muscular mass and more balanced structure~\cite{faria2009analise,weber2009parametros}. Although there is not an ideal biotype for all production systems, the adequate biotype should be determined according to the objectives that have been established for the herd, along with the production system being practiced~\cite{josahkian2013avaliacao}. This is not without consequences. For instance, larger animals usually have higher nutritional and general maintenance requirements~\cite{horimoto2006estimativas}.

Among the methods used to evaluate beef cattle, the EPMURAS methodology synthesized by~\citet{kouryfilho2005escores,kouryfilho2015avaliacao} is one of the most utilized in Brazil. It consists in a visual assessment of body structure, precocity, muscularity, sheath, racial aspects, angulation and sexuality. In this work, we investigate the structure scores, which should be highly correlated to the size of the animal. More specifically, the sources usually state that it is an assessment of the area of the animal as seen from the side, and that a grader should look at the length of the animal and at its rib height. The scores are investigated in their correlation with body measurements and productivity metrics. Ultimately, the objective is to mine scores data in order to propose a new way of classifying beef cattle, by application of the k-means clustering algorithm.

There are different systems to evaluate beef cattle, and, more specifically, to evaluate body structure. \citet{barbosa2006tamanho}, \citet{horimoto2006estimativas} and \citet{mercadante2007classificaccao}, for instance, investigated the applicability, in Brazil, of tables created by the U.S. Beef Improvement Federation (BIF), used for classifying structure \textit{qua} frame size in scores according to the croup height and age of the animals. The evaluation procedure differs from the one performed according to the EPMURAS methodology, even if the object under evaluation, the ``body structure'' in some sense, is the same.

In applying the EPMURAS methodology, subjectivity can be considered a problem. Ideally, as stated by \citet{kouryfilho2015avaliacao}, the largest 25\% of the animals belonging to the race should receive the highest score (6), while the smallest 25\% should score between 1 and 3, which means that a top-down approach could be applied to determine the actual measurements of these groups \textit{a posteriori}. In practice, however, the scores are attributed by a grader after simple visual inspection. In a study conducted with over 21 thousand nelore bulls, \citet{lima2013genetic} found that approximately 25\% of the animals did score between 1 and 3, but less than 10\% scored 6.

Another important measurement, specially for management, is the body weight of the animal. In the last years, many advancements have been made in order to improve the weighing, above all by using image-based techniques instead of proper physical scales, which can be stressful and harmful for the animals, as well as costly for the producers~\cite{zhao2023review}. Given the importance of body weight for management and decision making, it is also investigated alongside the structure scores.

The main contribution of this work, therefore, is the discussion of the correlation between body measurements and structure scores, along with the application of k-means clustering on measurements that correlate well with body weight and with structure scores. The results of the proposed study can be beneficial in some ways. First, it could improve structure scoring by making it less subjective, since the clusters will be based on measurements. Second, the classification proposed via clustering can be useful for management, by improving the separation of animals according to their structure or weight.




\section{Materials and Methods}
\label{methods}
\subsection{Data collection}

The data used in this study were obtained from 23 male Nelore cattle that participated in the Performance Evaluation Test (henceforth PET) performed by Embrapa Beef Cattle from March to June, 2021, in Campo Grande, Mato Grosso do Sul, Brazil. In general, the PET aims at evaluating feeding characteristics, such as intake and efficiency, as well as reproduction and carcass features of breeder animals classified as elite in genetic improvement programs. The animals came from different farms in Brazil, and were subjected to the same feeding, housing and handling conditions, which followed the procedure established by~\citet{gomes2013procedimenos}.

\begin{figure*}
     \centering
     \begin{subfigure}[b]{0.5\textwidth}
     \centering
         \includegraphics[width=0.9\textwidth,height=0.9\textwidth]{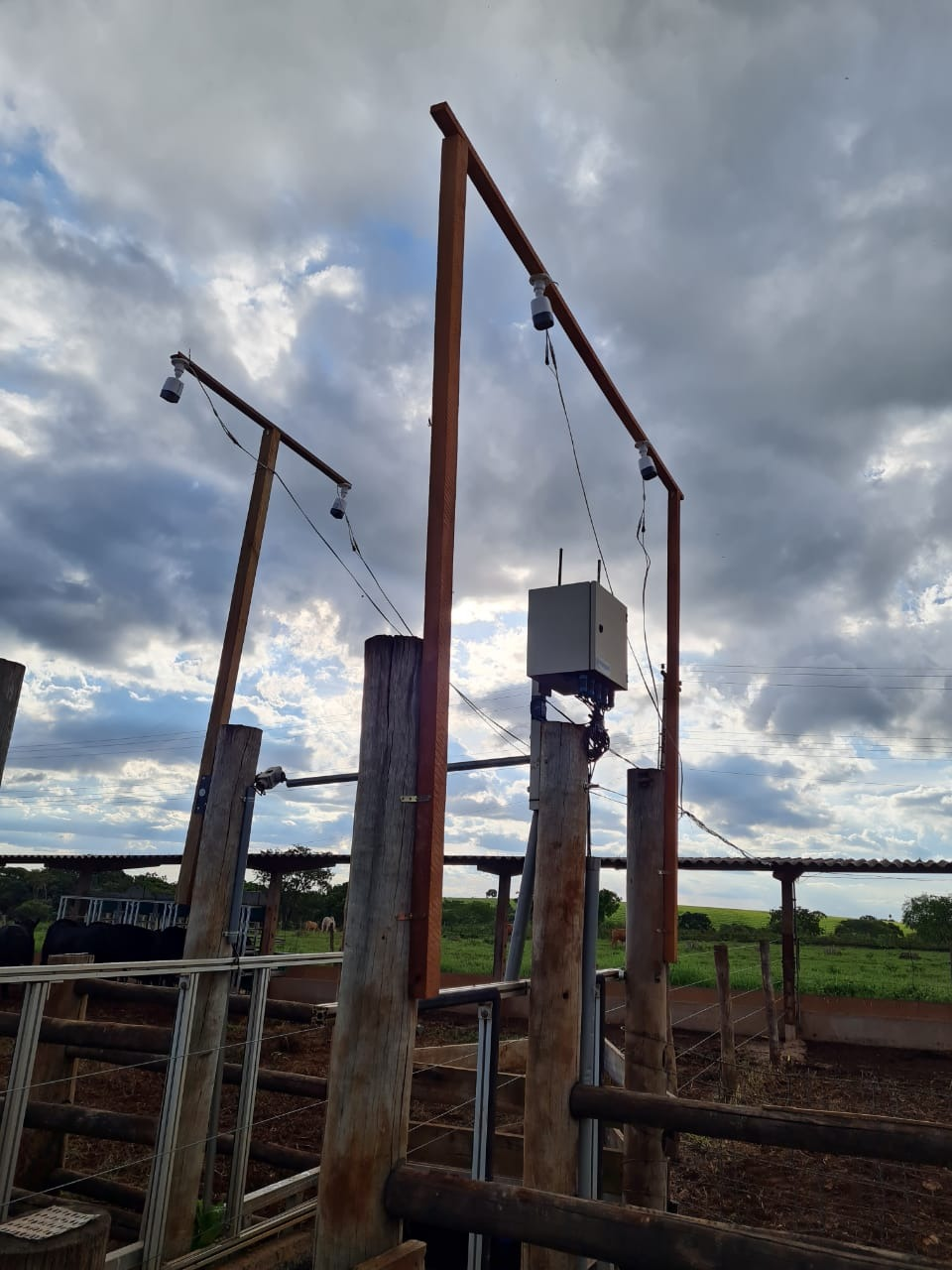}
         \label{fig:cameras_topo}
     \end{subfigure}%
     \begin{subfigure}[b]{0.5\textwidth}
         \centering
         \includegraphics[width=0.9\textwidth,height=0.9\textwidth]{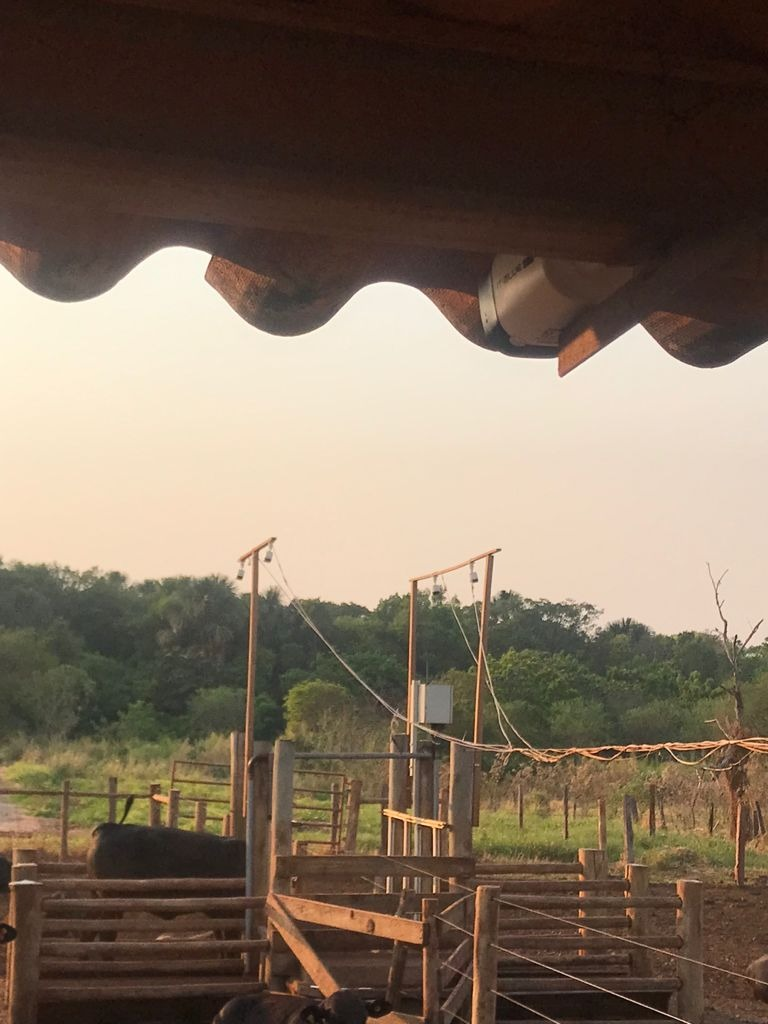}
         \label{fig:cameras_lateral}
     \end{subfigure}
     
     \caption{Equipment setup used for data gathering. Cameras were placed around a watering trough, in order to capture pictures both of the dorsum of the animals and also of their sides.}
     \label{fig:cameras}
\end{figure*}

For image acquisition, a DVR-MD-1004NS digital video recorder (MIDI JAPAN, China), with a recording capacity of 1 TB and an AHD 720p RGB camera, was used. The cameras were fixated on the metal structure of the trough's roof, in a way that allowed taking an image of the animal's body as seen from the side (profile picture), as well as of an image of the animal as seen from above (dorsum picture). Figure~\ref{fig:cameras} shows how the cameras were fixated. The videos were recorded at 30 frames per second. Subsequently, the videos were watched and frames were extracted, according to the following rule: whenever an animal accessed the watering trough, a frame was extracted if there was no obstruction by another animal and if the contour of the animal could be seen. This resulted in a different number of images for each animal, ranging from four up to seven pictures. Furthermore, at the moment of image collection, the animals had to go through a scale. Then, their body weight (BW) was measured. 

\begin{figure*}
     \centering
     \begin{subfigure}[b]{\textwidth}
     \centering
         \includegraphics[width=\textwidth]{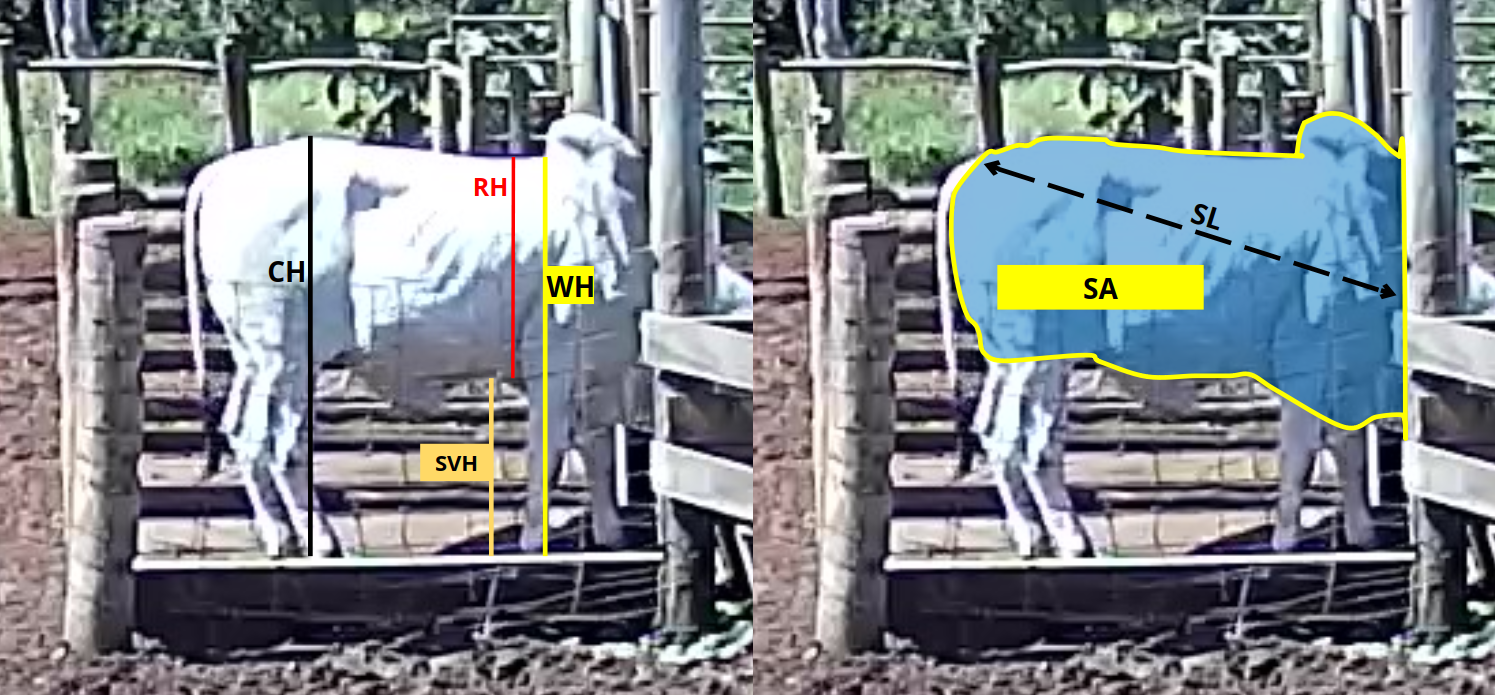}
         \caption{Measurements taken from images that show the side of the animal.}
         \label{fig:side_measurements}
     \end{subfigure}
     \begin{subfigure}[b]{\textwidth}
         \centering
         \includegraphics[width=\textwidth]{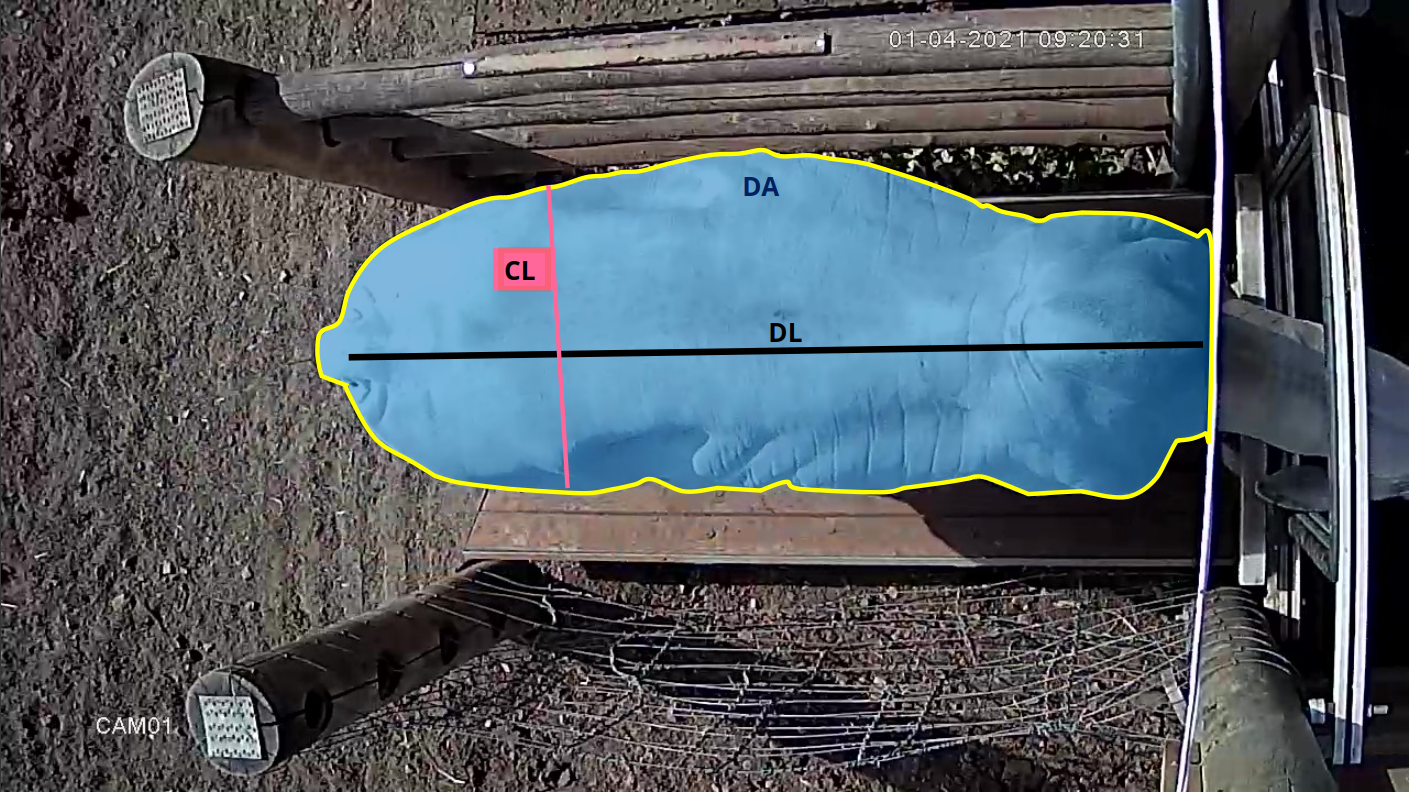}
         \caption{Measurements taken from images that show the animal from above. The set of these three measurements is being called ``dorsum view'' in this work.}
         \label{fig:top_measurements}
     \end{subfigure}
     
     \caption{Illustration of the measurements extracted from the images. For  measurement extraction, the ImageJ software was used. In order to convert the image measurements into the physical ones, parts of the structure of the trough were manually measured, and these measurements were used for calibration.}
     \label{fig:measurements}
\end{figure*}

In total, 116 profile pictures and 116 dorsum view pictures were selected (amounting to 232 images). After image collectionig, the ImageJ software~\cite{abramoff2004image} was used to extract the body measurements in Table~\ref{table:measurements}. Figure~\ref{fig:measurements} illustrates each of these measurements~\cite{weber2020prediction}. In order to convert the image-obtained measurement to the physical one, the structures around the watering trough were manually measured and used as reference. The extraction of the measurements was performed on all 232 images, which allowed the average of the measurements to be calculated. We opted for the average because, given the method used, some variation is expected, due to various factors, such as the animal's spatial positioning~\cite{mckiernan2005frame,rosa2014correlacoes,kamchen2021application}.

\newgeometry{margin=1cm}
\begin{landscape}
\begin{table*}[!htb]
    \centering
    \begin{tabular}{lcc}
        \toprule
        \multirow{2}{*}{\textbf{Measurement}} & \multirow{2}{*}{\textbf{Acronym}} & \multirow{2}{*}{\textbf{Origin}} \\ 
        &&\\
        \hline 
        Body weight & BW & Automatic weighing on animal approximation to the watering trough. \\ 
        \hline
        Croup height            & CH  & \multirow{9}{*}{Averaged over values obtained by manual extraction from images} \\ 
        Withers height          & WH  &    \\
        Rib height              & RH  &    \\
        Side length             & SL  &    \\
        Sub-sternal void height & SVH &    \\
        Side area               & SA  &    \\
        Croup width             & CW  &    \\
        Dorsum length           & DL  &    \\
        Dorsum area             & DA  &    \\ 
        \hline
        Final weight & FW* & \multirow{5}{*}{Productivity metrics obtained at the end of the PET} \\
        Dry matter intake & DMI* & \\
        Residual feed intake & RFI* & \\
        Average daily gain & ADG* & \\
        Scrotal circumference & SC* & \\
        \hline
        Loin eye area &  LEA* & Carcass ultrasound evaluation at the end of the PET \\
        \hline
        Scores given by the three graders & S1, S2, S3** & \multirow{2}{*}{The scores were attributed by specialists and then averaged} \\
        Structure scores & SS** &  \\
        \bottomrule
    \end{tabular}
    \caption{Different measurements were taken and utilized in this work. For those obtained from images, the ImageJ software was used, along with some manual measurements of the physical structure of the watering trough, which were used as reference for calibration. Whenever the pictures were taken, the animals were also weighed. Some productivity metrics were taken at the end of the PET. Also after the PET, the carcass was evaluated by ultrasound scanning methods. As a visual aid, these productivity metrics are indicated by a *. One random image of each animal was sent to human specialists for scoring. The scores are indicated by **.}
    \label{table:measurements}
\end{table*}

\end{landscape}
\restoregeometry

For each animal, one profile image was randomly selected and sent to three experienced specialists, who attributed structure scores following the so-called EPMURAS methodology~\cite{kouryfilho2005escores,kouryfilho2015avaliacao}. Structure scores are said to take into account both the rib height and the body length of the animal, and ranges from 1 up to 6. The three human specialists were experienced zootechnicians and graders. The first one had two years of experience with this kind of evaluation. The second one also had two years of experience with the application of the visual scoring system, and had also attended specialization courses. The third one had approximately one year of experience as grader, by the time of the evaluation. This shows that the scores utilized in this study are, inasmuch as the EPMURAS method is concerned, technically validated and reliable.

At the end of the PET, the animals were subjected to an ultrasonographic evaluation of the carcass, for the assessment of the loin area (LA), following the methodology of \citet{bonin2017utilizacao}. The ultrasound evaluations were performed by trained field technicians and followed the standards established by the Ultrasound Guidelines Council\footnote{\url{https://ultrasoundguidelinescouncil.org/}}. An Aloka SSD 500 Micrus ultrasound (Aloka Co. Ltd, Wallingford, CT) equipped with a 172 mm long linear transducer and a frequency of 3.5 MHz was used. The loin area measurements were obtained from cross-sectional images of the Longissimus muscle between the 12th and 13th rib. The ultrasound images were stored using an image capture system (UICS; Walter \& Associates, LLC, Ames, IA) and subsequently analyzed by certified technicians, responsible for the quality of the data.

At the end of the PET, the following productivity measurements were taken: final weight (FW), average daily gain (ADG), dry matter intake (DMI) and residual feed intake (RFI).

\subsection{Data analysis and clustering}
\label{clustering_methodology}

After image-based measurement, structure score attribution by human graders and computation of the average structure score, and acquisition of final PET results, descriptive statistics were calculated, along with Pearson's coefficient of correlation between all the measurements in Table~\ref{table:measurements}.

After analysing the correlations, two sets with three measurements each were selected for clustering, based on the strongest correlations with body weight and scores. The first set contains CW, DL and DA, since these measurements were found to correlate more strongly with the body weight. Furthermore, one should also point out that these measurements can be obtained from Figure~\ref{fig:top_measurements} alone, which could lead to the simplification of the system used for image collection, depending on the results. Besides this first group, which we shall henceforth call \textit{dorsum view}, a second group was chosen, comprising BW, CH and CW, which are the measurements that achieved the strongest (positive for BW and CW, negative for CH) correlation with the average structure scores. The downside of this group is, of course, that the measurements were taken from three different sources: two images (CH and CW) and the scale (BW).

For clustering, the data was first standardized by z-score. Then, the k-means clustering algorithm was utilized on each group of measurements selected as described above\footnote{In this study, all the necessary implementations were carried out in the Python programming language. The Scikit-learn~\cite{scikit-learn} implementation of the k-means algorithm and of the metrics was used. The elbow method, however, was applied with the implementation in the Yellowbrick package~\cite{bengfort_yellowbrick_2018}. The hypothesis tests, the Pingouing package was used~\cite{pingouin}. Finally, plots were generated with the Matplotlib~\cite{matplotlib} and Seaborn~\cite{seaborn} packages.}. To determine the ideal number of clusters (\textit{k}), the elbow method was used. Since clusters generated by k-means do not have an intrinsic order or hierarchy, after the clustering process, the clusters were sorted based on the first values of their centroids, since these are the values pertaining to the measurements that presented the highest correlations during the choice of measurements for clustering. However, one should notice that this is not a trivial choice, as we discuss in the next sections. In each case, clusters were assigned a number ranging from 1 to \textit{k}, where 1 should be the cluster comprising the worst animals and \textit{k} should comprise the best animals. In order to evaluate the results, Pearson's r was calculated between the clusters and the other measurements. Scatter plots and boxplots were also used for visual inspection. Finally, ANOVA was used to evaluate the clusters, taking BW and SS** as dependent variables, at 5\% significance threshold, along with Tukey's Honestly Significant Differences (TukeyHSD).

\section{Results and Discussion}

Descriptive statistics, including mean, standard deviation, first, second and third quantiles, and minimum and maximum values, for the measurements collected and utilized in this study are shown in Table~\ref{table:descriptive_stats}. On average, the animals weighed 565.11 ($\pm 77.82$) kg at the end of the PET. The average structure score observed in this study was 2.90, which is lower than that observed in works such as that by \citet{lima2013genetic} (4.19) and \citet{kouryfilho2009estimativas} (3,99). As discussed below, grader S3 may be biased towards a lower value. However, even if their grades are disregarded, the grades given by S1 and S2 were also, on average, lower than that found in those other studies.

\begin{table*}[htb!]
    \centering
    \caption{Descriptive statistics: mean, standard deviation, minimum values, quantiles and maximum values for the measurements obtained in this study.}
    \label{table:descriptive_stats}
    \begin{tabular}{lccccccc}
        \toprule
         & mean & std & min & 25\% & 50\% & 75\% & max \\
        \midrule
            BW & 422.85 & $\pm 36.18$ & 355.00 & 392.81 & 431.42 & 449.79 & 478.70 \\
            CH & 130.45 & $\pm 2.82$ & 125.79 & 128.67 & 130.72 & 132.01 & 135.40 \\
            WH & 122.07 & $\pm 2.64$ & 117.58 & 120.91 & 121.82 & 123.98 & 127.70 \\
            RH & 65.75 & $\pm 2.08$ & 61.17 & 64.55 & 65.39 & 67.05 & 69.58 \\
            SL & 136.47 & $\pm 3.98$ & 128.43 & 133.48 & 135.98 & 139.79 & 145.61 \\
            SVH & 55.91 & $\pm 3.00$ & 48.66 & 53.80 & 56.73 & 57.60 & 62.20 \\
            SA & 9718.98 & $\pm 586.53$ & 8610.04 & 9436.68 & 9745.94 & 10004.56 & 10884.05 \\
            CW & 50.12 & $\pm 2.29$ & 45.28 & 48.78 & 50.27 & 52.21 & 53.49 \\
            DL & 153.98 & $\pm 5.22$ & 142.55 & 151.02 & 156.19 & 157.36 & 161.67 \\
            DA & 7298.45 & $\pm 414.67$ & 6453.65 & 7010.30 & 7411.05 & 7588.86 & 7885.65 \\
            FW* & 565.11 & $\pm 77.82$ & 448.94 & 529.37 & 558.77 & 596.12 & 726.50 \\
            DMI* & 12.95 & $\pm 1.07$ & 11.32 & 12.09 & 12.87 & 13.78 & 15.38 \\
            RFI* & 0.18 & $\pm 0.69$ & -1.04 & -0.33 & -0.02 & 0.83 & 1.69 \\
            ADG* & 1.74 & $\pm 0.36$ & 1.03 & 1.52 & 1.69 & 1.94 & 2.67 \\
            SC* & 33.28 & $\pm 2.44$ & 29.50 & 31.65 & 32.90 & 35.25 & 38.16 \\
            LEA* & 80.54 & $\pm 9.99$ & 60.48 & 77.50 & 79.61 & 85.41 & 98.13 \\
            S1** & 3.04 & $\pm 1.22$ & 2.00 & 2.00 & 3.00 & 4.00 & 5.00 \\
            S2** & 3.09 & $\pm 1.35$ & 1.00 & 2.50 & 3.00 & 4.00 & 6.00 \\
            S3** & 2.57 & $\pm 1.24$ & 1.00 & 1.50 & 3.00 & 3.00 & 5.00 \\
            SS** & 2.90 & $\pm 1.17$ & 1.33 & 2.17 & 2.67 & 3.50 & 5.33 \\
        \bottomrule
    \end{tabular}
\end{table*}

Table~\ref{table:scores} shows the scores given by human specialists for each animal, along with their average values and the standard deviations between graders. As one can see, animal 4 achieved the highest average (5.3). The scores for animal 5 had the highest standard deviation ($\pm 1.15$), but two out of three evaluators agreed on score 5. On the opposite side, all graders agreed on the score of animal 14 (3). Finally, in three cases, animals 10, 18 and 19, the three grades disagreed, but one should notice that the three grades were always adjacent to each other (STD $ = 1.00$).

The closeness of the scores even in cases of disagreement indicates that, although the graders weighed specific aspects differently, there was an overall agreement as to the quality of the cattle regarding its structure. One could argue that animal 5 was an exception. From Table~\ref{table:descriptive_stats}, it is possible to see that the grades given by grader S3 were, on average, considerably lower than those given by S1 and S2. In fact, in Table~\ref{table:scores} it is also possible to see that, whenever all graders disagreed, S3 gave the lowest score. This allows the conjecture that some kind of bias is at play here, and therefore that the grades given by S3 were lower than they should be. Both the agreement and the possible bias are to be expected, given the fact that EPMURAS is an evaluation, and that it is, nonetheless, influenced by subjective factors.

\begin{table}[htb!]
    \centering
    \begin{tabular}{lccccc}
        \toprule
         Animal & S1 & S2 & S3 & Mean (STD) \\
        \midrule
        1 & 2 & 3 & 2 & 2.3 ($\pm 0.58$) \\
        2 & 2 & 1 & 1 & 1.3 ($\pm 0.58$) \\
        3 & 5 & 5 & 4 & 4.7 ($\pm 0.58$) \\
        4 & 5 & 6 & 5 & 5.3 ($\pm 0.58$) \\
        5 & 5 & 5 & 3 & 4.3 ($\pm 1.15$) \\
        6 & 4 & 3 & 3 & 3.3 ($\pm 0.58$) \\
        7 & 5 & 4 & 5 & 4.7 ($\pm 0.58$) \\
        8 & 3 & 4 & 3 & 3.3 ($\pm 0.58$) \\
        9 & 5 & 4 & 4 & 4.3 ($\pm 0.58$) \\
        10 & 3 & 4 & 2 & 3.0 ($\pm 1.00$) \\
        11 & 2 & 3 & 2 & 2.3 ($\pm 0.58$) \\
        12 & 2 & 2 & 3 & 2.3 ($\pm 0.58$) \\
        13 & 2 & 1 & 1 & 1.3 ($\pm 0.58$) \\
        14 & 3 & 3 & 3 & 3.0 ($\pm 0.00$) \\
        15 & 2 & 3 & 3 & 2.7 ($\pm 0.58$) \\
        16 & 2 & 1 & 1 & 1.3 ($\pm 0.58$) \\
        17 & 2 & 1 & 1 & 1.3 ($\pm 0.58$) \\
        18 & 3 & 2 & 1 & 2.0 ($\pm 1.00$) \\
        19 & 2 & 3 & 1 & 2.0 ($\pm 1.00$) \\
        20 & 2 & 3 & 3 & 2.7 ($\pm 0.58$) \\
        21 & 3 & 3 & 2 & 2.7 ($\pm 0.58$) \\
        22 & 2 & 3 & 3 & 2.7 ($\pm 0.58$) \\
        23 & 4 & 4 & 3 & 3.7 ($\pm 0.58$) \\
        \bottomrule
    \end{tabular}
    \caption{Structure scores for each animal. One image of each animal was randomly selected and sent to three specialized technicians, who graded them for structure, according to the EPMURAS methodology.}
    \label{table:scores}
\end{table}

\begin{figure*}[htb!]
    \centering
    \includegraphics[width=\textwidth]{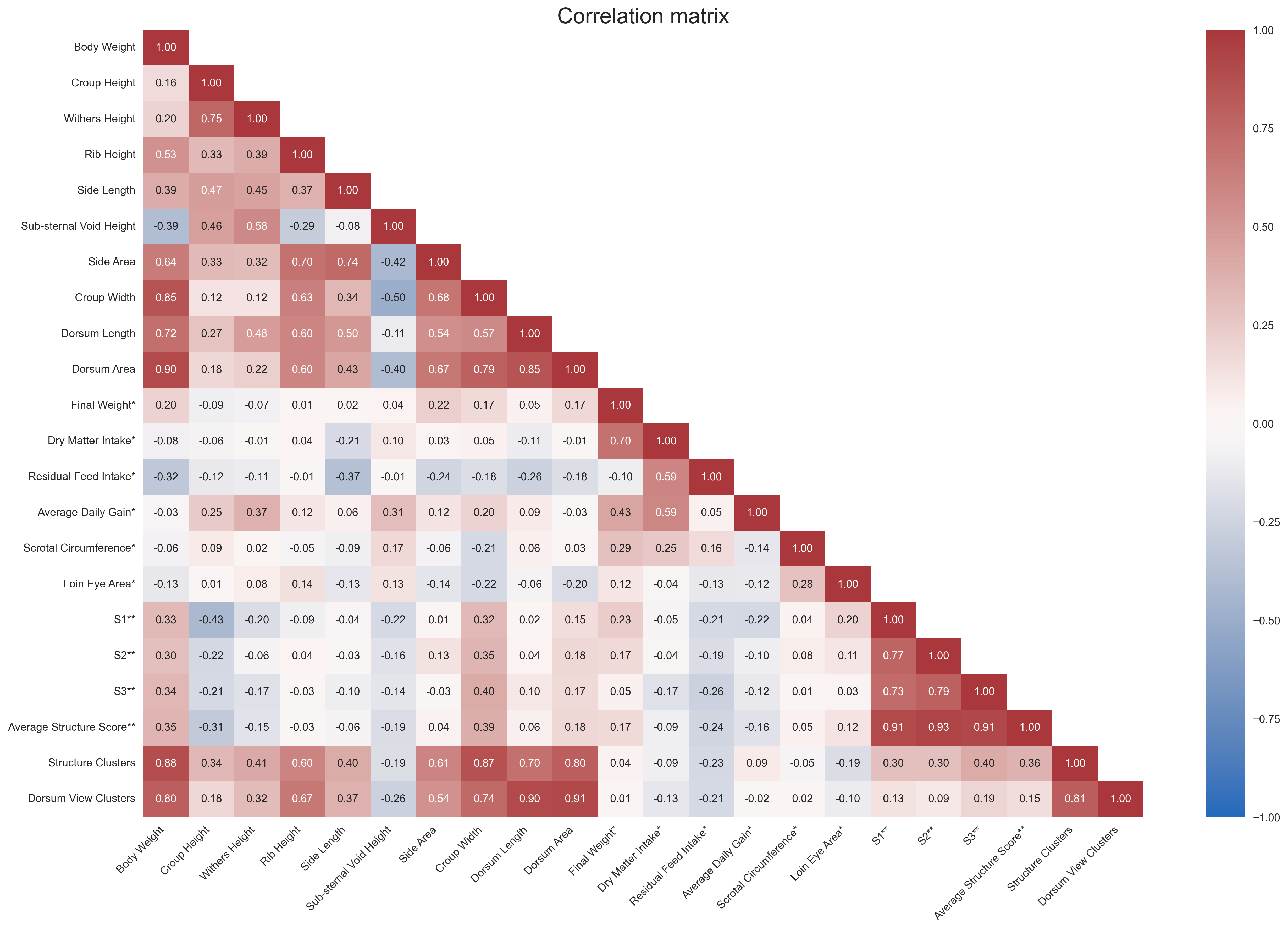}
    \caption{Correlation matrix for the measurements utilized in this study.}
    \label{fig:corr_matrix}
\end{figure*}

Figure~\ref{fig:corr_matrix} shows the correlation matrix between the body weight, the measurements taken from the images, the productivity metrics calculated after the PET, and the structure scores\footnote{For simplicity, only one correlation matrix is shown in this work, in Figure~\ref{fig:corr_matrix}. This matrix comprises the correlations between every variable analysed throughout the study. However, the reader should keep in mind, for a clear understanding of the procedure, that the correlations between the variables in Table~\ref{table:measurements} were calculated \textit{before} clustering, whereas the correlations that involve clusters were calculated only \textit{after} the clusters were created.}. It also shows the correlations between the clusters and the other variables used in this study. Some strong (or almost strong) positive correlations can be observed between certain measurements, mainly involving the weight: DA (0.9) and CW (0.85). It is also possible to notice that there is not a strong correlation between the productivity metrics and the image measurements. In this case, the strongest correlations that were observed are 0.37 between ADG* and WH, and -0.37 between RFI* and SL. Between two metrics, the strongest correlation was that between DMI* and FW* (0.70).

\begin{figure}[!htb]
     \centering
     \begin{subfigure}[b]{0.5\textwidth}
         \centering
         \includegraphics[width=\columnwidth]{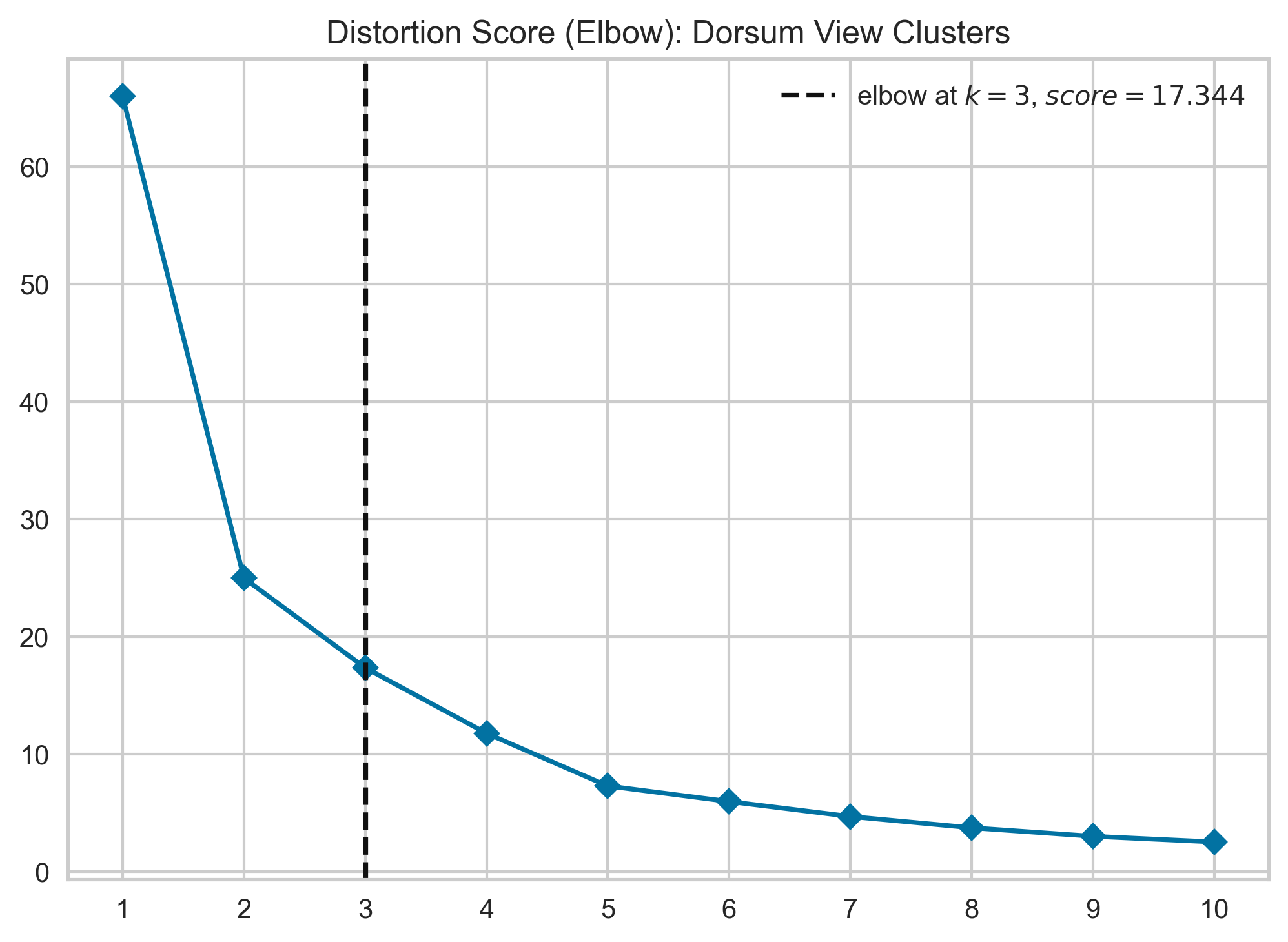}
         \caption{}
         \label{fig:dorsum_view_elbow}
     \end{subfigure}%
     \hfill
     \begin{subfigure}[b]{0.5\textwidth}
         \centering
         \includegraphics[width=\columnwidth]{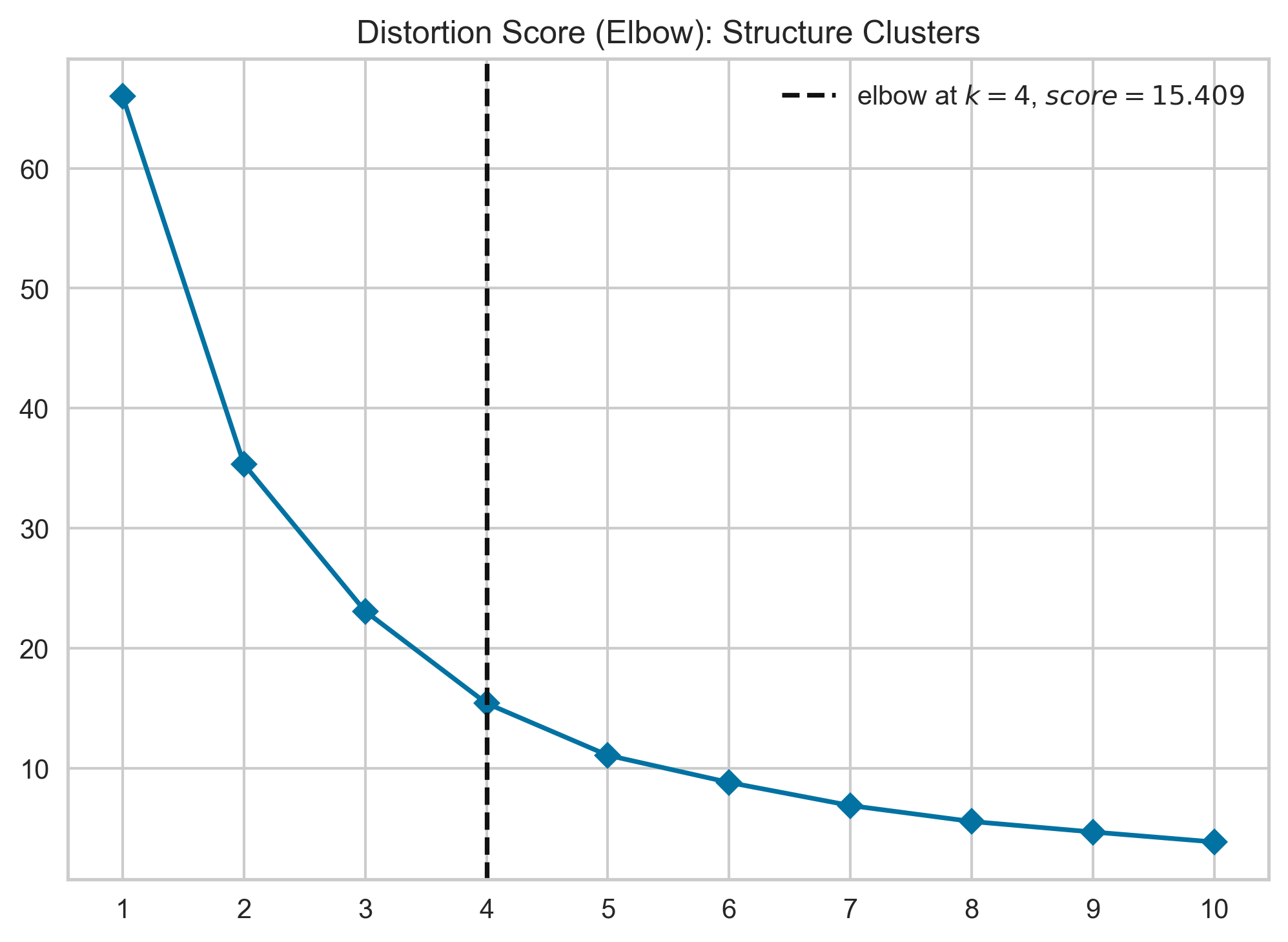}
         \caption{}
         \label{fig:dorsum_view_normalized_elbow}
     \end{subfigure}%
        \caption{Distortion scores calculated in the application of the elbow method for finding the optimum value of $k$. Evaluated values were in the range $[1, 10]$. For the dorsum view, the method yielded $k=3$ as the optimum value, whereas it yielded $k=4$ as the optimum value for the structure clusters. Z-score normalization was applied prior to the use of the elbow method.}
        \label{fig:elbow}
\end{figure}

\begin{figure*}[!htb]
     \centering
     \begin{subfigure}[b]{0.5\textwidth}
         \centering
         \includegraphics[width=\columnwidth]{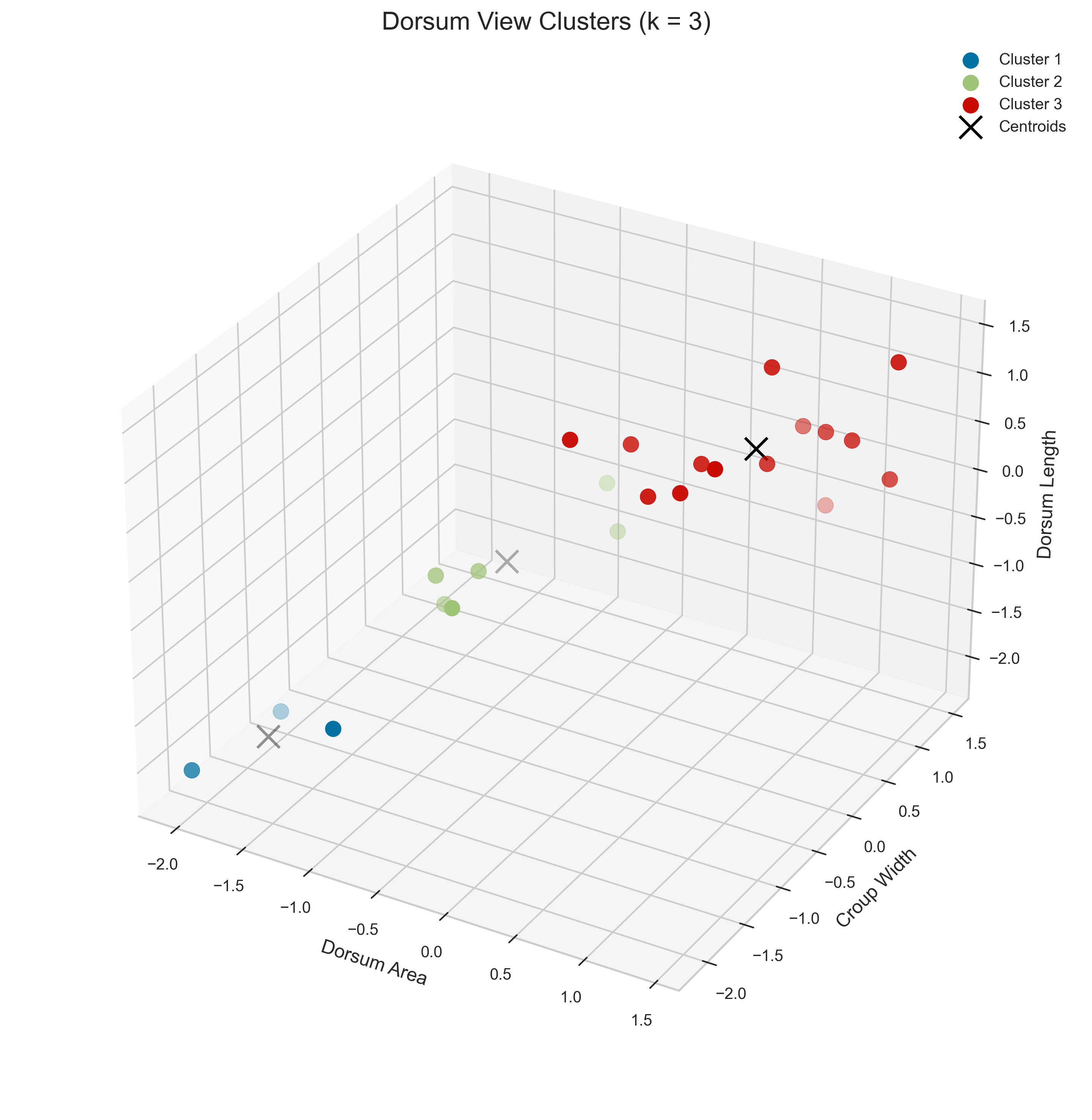}
         \caption{}
         \label{fig:clusters_dorsum_view}
     \end{subfigure}%
     \hfill
     \begin{subfigure}[b]{0.5\textwidth}
         \centering
         \includegraphics[width=\columnwidth]{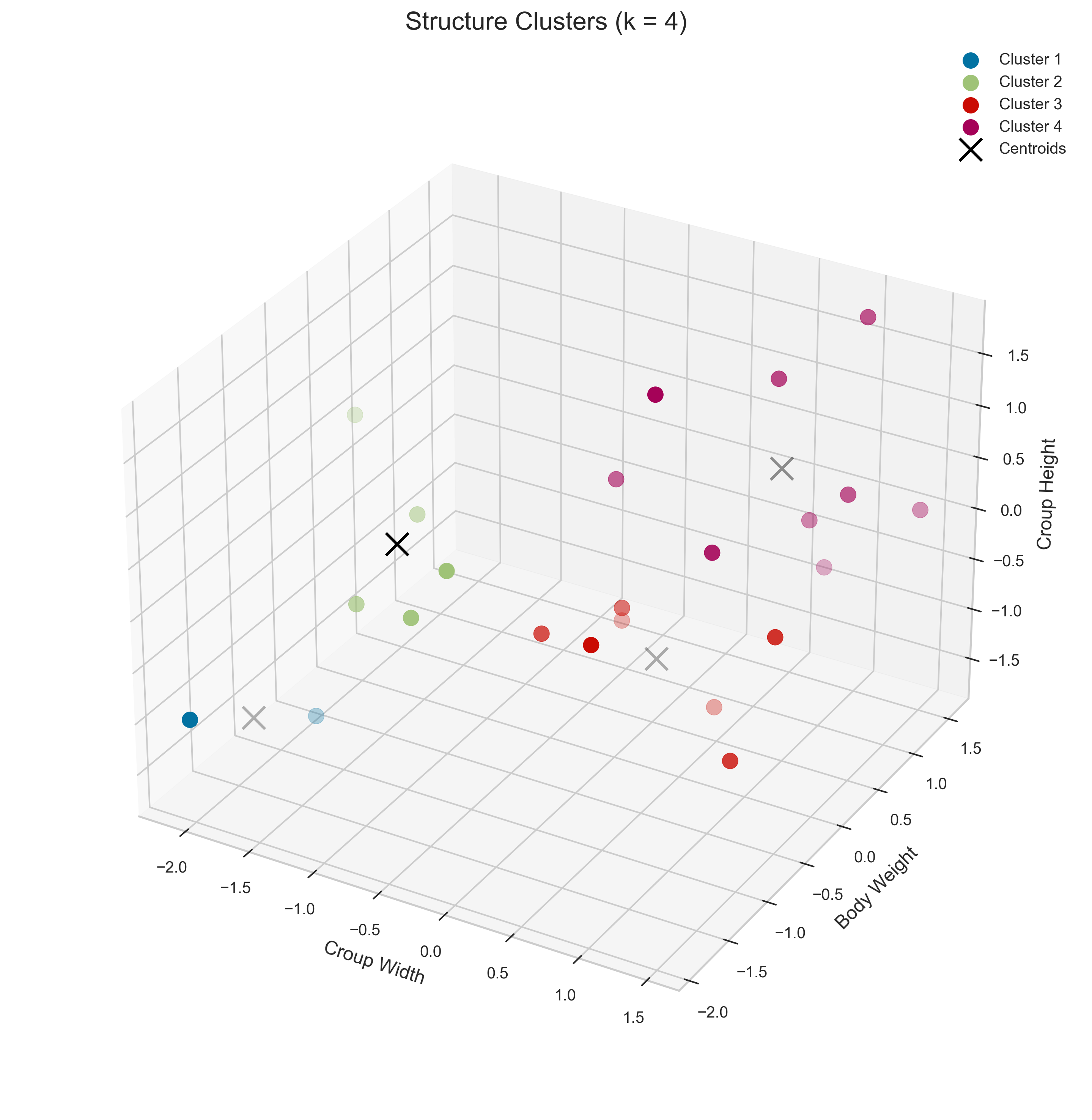}
         \caption{}
         \label{fig:clusters_structure}
     \end{subfigure}%
        \caption{Clusters generated by k-means, along with their centroids. Z-score normalization was applied on the values, before the usage of the k-means algorithm.}
        \label{fig:clusters}
\end{figure*}

\begin{figure}[!htb]
     \centering
     \begin{subfigure}[b]{0.5\textwidth}
         \centering
         \includegraphics[width=\columnwidth]{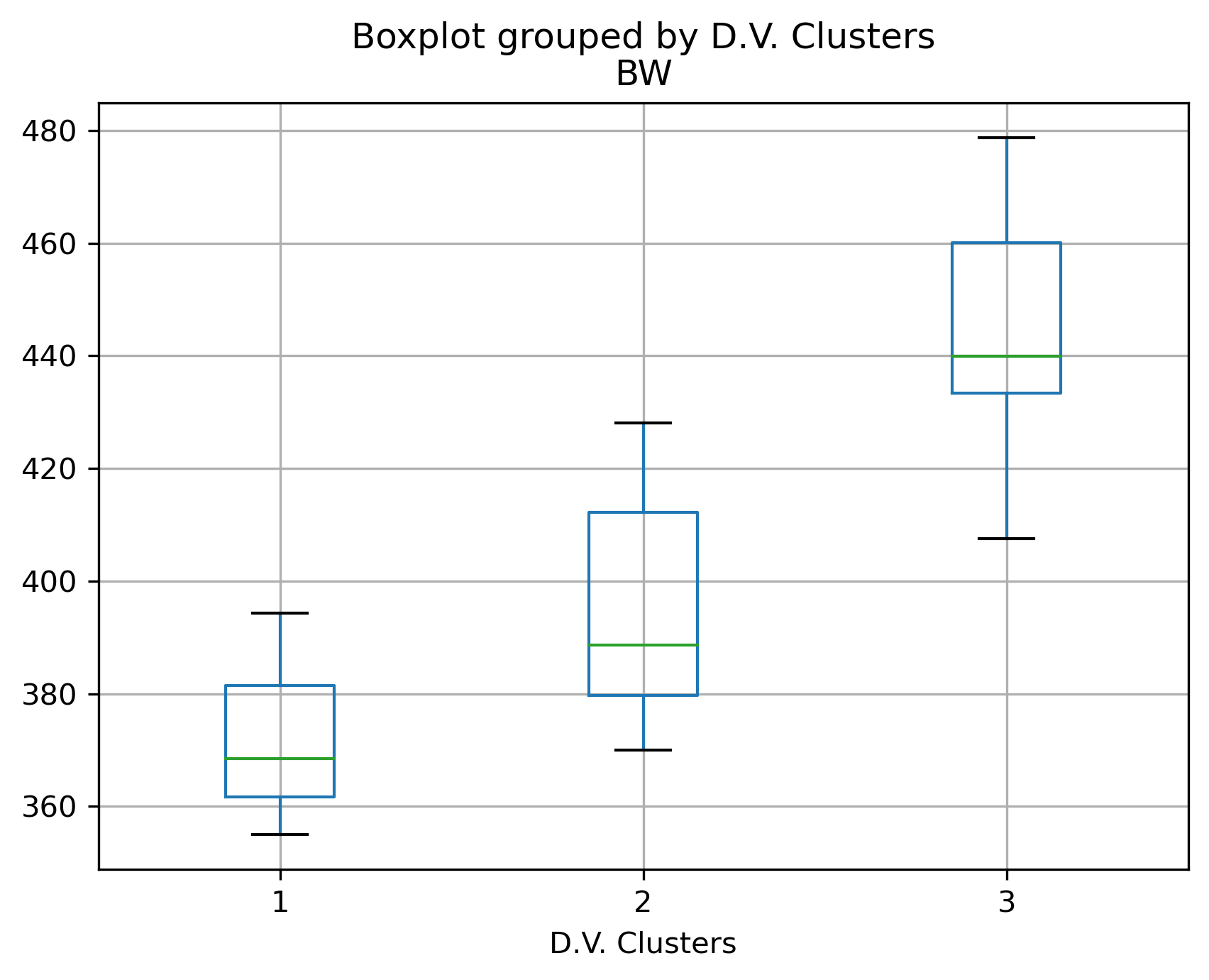}
         \caption{}
         \label{fig:boxplot_dorsum_view}
     \end{subfigure}%
     \hfill
     \begin{subfigure}[b]{0.5\textwidth}
         \centering
         \includegraphics[width=\columnwidth]{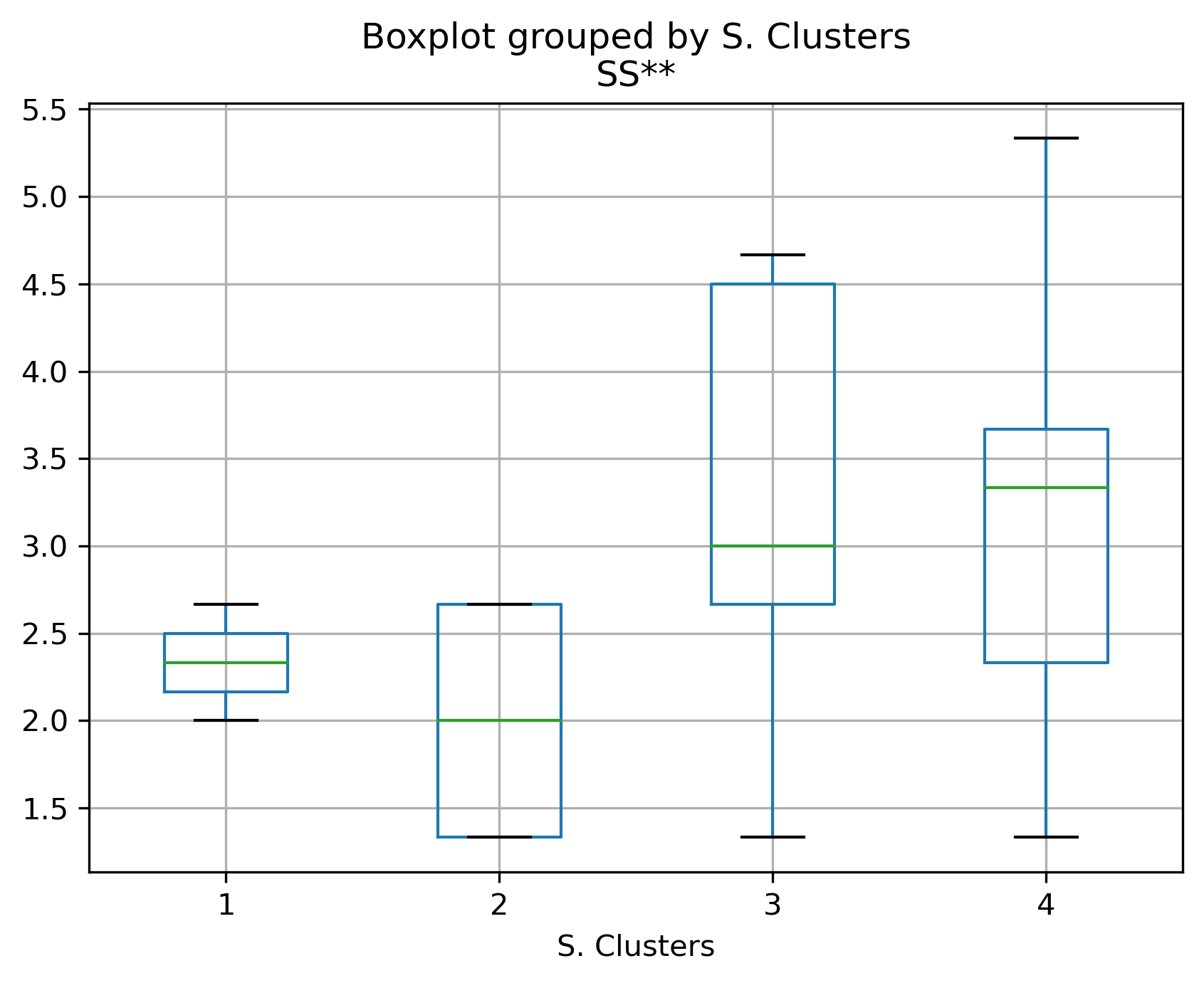}
         \caption{}
         \label{fig:boxplot_structure}
     \end{subfigure}%
        \caption{Boxplots for the k-means results. One can see that visually the dorsum view generated clusters that separated well the BW of the animals. On the other hand, the variables chosen for their correlation with structure did not achieve a good separation, when the average score of the groups is considered. In any case, one should notice that the ANOVA did was not significant for the structure clusters. For the dorsum view, TukeyHSD indicated that clusters 1 and 2 did not differ amongst themselves, but both differed from cluster 3.}
        \label{fig:boxplots}
\end{figure}

Since DA, CW and DL achieved the strongest observed correlation with the body weight (0.9, 0.85 and 0.72), they were chosen as input variables for the k-means, as describe in Section~\ref{clustering_methodology}. These correlations are in agreement with the existing literature. \citet{weber2020prediction}, for instance, also found that dorsum area and croup width are two measurements strongly correlated with the body weight.

Only weak positive correlations were found between the structure scores and the image-obtained measurements. Regarding these, two observations are in order. First, since CW, BW and CH achieved the strongest observed correlations with the average structure scores (respectively, 0.39, 0.35 and -0.31), they were also chosen for the application of the k-means clustering, as described in Section~\ref{clustering_methodology}, even if this correlation is not strong.

Second, we should address the fact that these weak correlations are at odds with the existing literature. According to the existing literature on the EPMURAS methodology~\cite{kouryfilho2005escores,kouryfilho2015avaliacao,josahkian2013avaliacao}, structure scores, inasmuch as they are an evaluation of the area of the animal as seen from the side, were expected to be strongly correlated with measurements such as the side area, the side length and the rib height of the animals. A possible reason for the weak correlation that was observed is that there may be an issue with the graders (\textit{e.g.}, bias, improper knowledge, etc.). However, this is unlikely, since the correlations were not inconsistent between the graders, as discussed above. One could also point out that the correlation between S1 and the CH was observed to be -0.43 against -0.22 and -0.21 for S2 and S3, respectively. Even then, in all cases the correlation would still be weak (considering $\pm 0.5$ as the threshold for moderate correlation, either positive or negative). Another possible reason for the discrepancy between our results and what was to be expected in virtue of the concepts would be that, since the grading procedure as carried out is influenced by subjective factors, the actual measurements evaluated by the graders can be more subtle and complex than those extracted by us.

Finally, one could be tempted to ask whether these results do not indicate that there is an issue with the EPMURAS method itself. While there may be a problem with the way the ``structure'' has been conceptualized (while the conceptualizations are usually not rigorously formally expressed, there are many illustrations that make it quite clear what is tended to during an evaluation of body structure - one can see, for instance, the work by~\citet{lima2012parametros}, as well as the seminal work by~\citet{kouryfilho2005escores}), the consistency between graders also indicates that the scores are not arbitrary. To the best of our knowledge, this aspect does not seem to have been investigated in depth. More data should be gathered for any conclusive statement, and this is, therefore, left for future work.

Figure~\ref{fig:elbow} shows the results yielded by the elbow method for finding the optimum number of clusters \footnote{as stated in Section~\ref{clustering_methodology}, z-score normalization was applied on all attributes. The results shown in this Section are presented with the normalization.}. For the dorsum view, the method yielded $k=3$, whereas for the structure the method resulted in $k=4$. Figure~\ref{fig:clusters} shows the two clustering results obtained according to the procedure describe in Section~\ref{clustering_methodology}. Figure~\ref{fig:boxplots} shows boxplots where the values for BW and SS** were grouped according to the clusters. The ANOVA results did not suggest any difference between the four structure clusters ($p=0.17$). The result was significant for the dorsum view ($p=0.000017$). The TukeyHSD for the dorsum view suggested that there was no difference between clusters 1 and 2 ($p=0.33$), but it was significant for both 1 and 3 ($p=0.00012$), and 2 and 3 ($p=0.00040$).

Truly, the choice of $k$ by means of the elbow should not be considered uncontroversial. By looking at the results of the elbow, one could ask whether $k=4$ or even $k=5$ could not be better for any of the groupings. By looking at the boxplots, and by recalling a little bit of domain knowledge and usages, one could also suggest that $k=3$ could be a good choice, since this would mimic the traditional and usual categories, such as ``head'', ``middle'' and ``back'', or even ``big'', ``medium'' and ``small'' cattle. A deeper investigation on the ideal \textit{k} for the groupings is left for future work.

From Figure~\ref{fig:clusters}, it is possible to argue that the clusters generated by the dorsum view contain points that are closer to each other. On the other hand, it is also possible to see that there is some confusion between clusters 2 and 3, with two points of cluster 2 being close to the points in cluster 3. Also, one can see that the number of points in the clusters is uneven, which may have contributed for the results yielded by the ANOVA. Visually, the structure clusters seem to be better separated, but the points in each cluster spread more across the space. Finally, the unevenness in the number of points per cluster remains - cluster 1 comprises only two data points. While we stick with $k=4$, this last observation could be a hint for adopting a different $k$ for the structure clusters, in addition to the possibilities already discussed above.

In Figure~\ref{fig:clusters}, the centroid of each cluster is indicated by an X. The specific coordinates of each cluster can be seen in Table~\ref{table:centroids}. From Table~\ref{table:centroids}, it is possible to say that the dorsum view clusters resulted in a quite straightforward categorization. Animals with higher dorsum view cluster score are animals whose dorsum is larger (regarding area) and longer, and whose croup is wider. On the other hand, the results of the structure clusters were not as straightforward. CW and BW increased together (which is expected, since they are almost strongly correlated between themselves (0.85)) and presented an almost strong correlation with the clusters (0.88 for BW, 0.87 for CW). However, CH ended up being only weakly correlated with the clusters (0.34). Finally, none of the clusters generated in this study led to more than a weak correlation with the productivity metrics.

\begin{table}[!htb]
    \centering
    \begin{tabular}{cccc}
    \hline
    \multicolumn{4}{c}{Dorsum View} \\
    \hline
    Cluster & Dorsum Area & Croup Width & Dorsum Length \\
    \hline
    1 & -1.72 & -1.67 & -1.87 \\
    2 & -0.71 & -0.32 & -0.55 \\
    3 & 0.67 & 0.49 & 0.64 \\
    \hline \hline
    \multicolumn{4}{c}{Structure Clusters} \\
    \hline
    Cluster & Croup Width & Body Weight & Croup Height \\
    \hline 
    1 & -1.98 & -1.33 & -1.34 \\ 
    2 & -0.90 & -1.22 & 0.68 \\
    3 & 0.26 & 0.11 & -0.90 \\
    4 & 0.74 & 0.89 & 0.62 \\
    \hline 
    \end{tabular}
    \caption{Coordinates of the centroids of each cluster generated by k-means. The reader should keep in mind that z-score standardization was applied on the values.}
    \label{table:centroids}
\end{table}

\section{Conclusion}

In this study, we collected, presented, investigated and clusterized data on body measurements, structure scores and productivity metrics of nelore cattle. While the results were not encouraging, and while it is still uncertain whether or not the new clusters can be useful for nelore cattle management (they are most certainly not useful for the prediction of productivity), we hope to have shown, as of the discussion, that the problem is interesting enough to be studied further. While the EPMURAS methodology is widely used, it was not clear, from the data collected and analysed in this study, what is actually observed when structure scores are attributed.

Furthermore, and finally, many points are left open for future research, which will require more data collection. Among them, one can mention: refine the protocol for application of clustering algorithms, with the substitution of the elbow method for more critically examined criteria to determine the number of clusters, as well as to determine the order imposed on the clusters \textit{a posteriori}; an in-depth investigation of the correlation between structure scores and visual aspects, potentially with the extraction of new features from the images; and a better use of structure scores for clustering, which shall stem from the aforementioned study.

\section{Acknowledgments}
This work has received financial support from the Dom Bosco Catholic University and the Foundation for the Support and Development of Education, Science and Technology from the State of Mato Grosso do Sul, FUNDECT. Some of the authors have been awarded with Scholarships from the the Brazilian National Council of Technological and Scientific Development, CNPq and the Coordination for the Improvement of Higher Education Personnel, CAPES. We would also like to thank NVIDIA for providing the Titan X GPUs used in the experiments.






\end{document}